
%
%
%

\input psfig



%
%
%
%
%


\hsize=6.0truein
\vsize=8.5truein
\voffset=0.25truein
\hoffset=0.1875truein
\tolerance=1000
\hyphenpenalty=500
\def\monthintext{\ifcase\month\or January\or February\or
   March\or April\or May\or June\or July\or August\or
   September\or October\or November\or December\fi}


\font\tenrm=cmr10 scaled \magstep1   \font\tenbf=cmbx10 scaled \magstep1
\font\sevenrm=cmr7 scaled \magstep1  
\font\fiverm=cmr5 scaled \magstep1   

\font\teni=cmmi10 scaled \magstep1   \font\tensy=cmsy10 scaled \magstep1
\font\seveni=cmmi7 scaled \magstep1  \font\sevensy=cmsy7 scaled \magstep1
\font\fivei=cmmi5 scaled \magstep1   \font\fivesy=cmsy5 scaled \magstep1

\font\tentt=cmtt10 scaled \magstep1
\font\tenit=cmti10 scaled \magstep1
\font\tensl=cmsl10 scaled \magstep1

\def\twelvepoint{\def\rm{\fam0\tenrm}
   \textfont0=\tenrm \scriptfont0=\sevenrm \scriptscriptfont0=\fiverm
   \textfont1=\teni  \scriptfont1=\seveni  \scriptscriptfont1=\fivei
   \textfont2=\tensy \scriptfont2=\sevensy \scriptscriptfont2=\fivesy
   \textfont\itfam=\tenit \def\it{\fam\itfam\tenit}
   \textfont\ttfam=\tentt \def\tt{\fam\ttfam\tentt}
   \textfont\bffam=\tenbf \def\bf{\fam\bffam\tenbf}
   \textfont\slfam=\tensl \def\sl{\fam\slfam\tensl} \rm
   \hfuzz=1pt\vfuzz=1pt
   \setbox\strutbox=\hbox{\vrule height 10.2pt depth 4.2pt width 0pt}
   \parindent=24pt\parskip=1.2pt plus 1.2pt
   \topskip=12pt\maxdepth=4.8pt\jot=3.6pt
   \normalbaselineskip=14.4pt\normallineskip=1.2pt
   \normallineskiplimit=0pt\normalbaselines
   \abovedisplayskip=13pt plus 3.6pt minus 5.8pt
   \belowdisplayskip=13pt plus 3.6pt minus 5.8pt
   \abovedisplayshortskip=-1.4pt plus 3.6pt
   \belowdisplayshortskip=13pt plus 3.6pt minus 3.6pt
   \topskip=12pt \splittopskip=12pt
   \scriptspace=0.6pt\nulldelimiterspace=1.44pt\delimitershortfall=6pt
   \thinmuskip=3.6mu\medmuskip=3.6mu plus 1.2mu minus 1.2mu
   \thickmuskip=4mu plus 2mu minus 1mu
   \smallskipamount=3.6pt plus 1.2pt minus 1.2pt
   \medskipamount=7.2pt plus 2.4pt minus 2.4pt
   \bigskipamount=14.4pt plus 4.8pt minus 4.8pt}

\twelvepoint



\font\titlerm=cmr10 scaled \magstep3
\font\titlerms=cmr10 scaled \magstep1 
\font\titlei=cmmi10 scaled \magstep3  
\font\titleis=cmmi10 scaled \magstep1 
\font\titlesy=cmsy10 scaled \magstep3 	
\font\titlesys=cmsy10 scaled \magstep1  
\font\titleit=cmti10 scaled \magstep3	
\skewchar\titlei='177 \skewchar\titleis='177 
\skewchar\titlesy='60 \skewchar\titlesys='60 

\def\titlefont{\def\rm{\fam0\titlerm}
   \textfont0=\titlerm \scriptfont0=\titlerms 
   \textfont1=\titlei  \scriptfont1=\titleis  
   \textfont2=\titlesy \scriptfont2=\titlesys 
   \textfont\itfam=\titleit \def\it{\fam\itfam\titleit} \rm}


\def\preprint#1{\baselineskip=19pt plus 0.2pt minus 0.2pt \pageno=0
   \begingroup
   \nopagenumbers\parindent=0pt\baselineskip=14.4pt\rightline{#1}}
\def\title#1{
   \vskip 0.9in plus 0.45in
   \centerline{\titlefont #1}}
\def\secondtitle#1{}
\def\author#1#2#3{\vskip 0.9in plus 0.45in
   \centerline{{\bf #1}\myfoot{#2}{#3}}\vskip 0.12in plus 0.02in}
\def\secondauthor#1#2#3{}
\def\addressline#1{\centerline{#1}}
\def\abstract{\vskip 0.7in plus 0.35in
	\centerline{\bf Abstract}
	\smallskip}
\def\finishtitlepage#1{\vskip 0.8in plus 0.4in
   \leftline{#1}\supereject\endgroup}

\def\date#1{\finishtitlepage{#1}}

\def\nolabels{\def\eqnlabel##1{}\def\eqlabel##1{}\def\figlabel##1{}%
	\def\reflabel##1{}}
\def\writelabels{\def\eqnlabel##1{%
	{\escapechar=` \hfill\rlap{\hskip.11in\string##1}}}%
	\def\eqlabel##1{{\escapechar=` \rlap{\hskip.11in\string##1}}}%
	\def\figlabel##1{\noexpand\llap{\string\string\string##1\hskip.66in}}%
	\def\reflabel##1{\noexpand\llap{\string\string\string##1\hskip.37in}}}
\nolabels


\global\newcount\secno \global\secno=0
\global\newcount\meqno \global\meqno=1
\global\newcount\subsecno \global\subsecno=0

\font\secfont=cmbx12 scaled\magstep1

\def\section#1{\global\advance\secno by1
   \xdef\secsym{\the\secno.}
   \global\subsecno=0
   \global\meqno=1\bigbreak\medskip
   \noindent{\secfont\the\secno. #1}\par\nobreak\smallskip\nobreak\noindent}

\def\subsection#1{\global\advance\subsecno by1
\medskip
\noindent
{\bf\the\secno.\the\subsecno\ #1}
\par\medskip\nobreak\noindent}

\def\newsec#1{\global\advance\secno by1
   \xdef\secsym{\the\secno.}
   \global\meqno=1\bigbreak\medskip
   \noindent{\bf\the\secno. #1}\par\nobreak\smallskip\nobreak\noindent}
\xdef\secsym{}

\def\appendix#1#2{\global\meqno=1\xdef\secsym{\hbox{#1.}}\bigbreak\medskip
\noindent{\bf Appendix #1. #2}\par\nobreak\smallskip\nobreak\noindent}

\def\acknowledgements{\bigbreak\medskip\centerline{\bf
   Acknowledgements}\par\nobreak\smallskip\nobreak\noindent}


\def\eqnn#1{\xdef #1{(\secsym\the\meqno)}%
	\global\advance\meqno by1\eqnlabel#1}
\def\eqna#1{\xdef #1##1{\hbox{$(\secsym\the\meqno##1)$}}%
	\global\advance\meqno by1\eqnlabel{#1$\{\}$}}
\def\eqn#1#2{\xdef #1{(\secsym\the\meqno)}\global\advance\meqno by1%
	$$#2\eqno#1\eqlabel#1$$}


\def\myfoot#1#2{{\baselineskip=14.4pt plus 0.3pt\footnote{#1}{#2}}}
\global\newcount\ftno \global\ftno=1
\def\foot#1{{\baselineskip=14.4pt plus 0.3pt\footnote{$^{\the\ftno}$}{#1}}%
	\global\advance\ftno by1}


\global\newcount\refno \global\refno=1
\newwrite\rfile

\def\ref{[\the\refno]\nref}
\def\nref#1{\xdef#1{[\the\refno]}\ifnum\refno=1\immediate
	\openout\rfile=refs.tmp\fi\global\advance\refno by1\chardef\wfile=\rfile
	\immediate\write\rfile{\noexpand\item{#1\ }\reflabel{#1}\pctsign}\findarg}
\def\findarg#1#{\begingroup\obeylines\newlinechar=`\^^M\passarg}
	{\obeylines\gdef\passarg#1{\writeline\relax #1^^M\hbox{}^^M}%
	\gdef\writeline#1^^M{\expandafter\toks0\expandafter{\striprelax #1}%
\edef\next{\the\toks0}\ifx\next\null\let\next=\endgroup\else\ifx\next\empty%

\else\immediate\write\wfile{\the\toks0}\fi\let\next=\writeline\fi\next\relax}}
	{\catcode`\%=12\xdef\pctsign{
\def\striprelax#1{}

\def\semi{;\hfil\break}
\def\addref#1{\immediate\write\rfile{\noexpand\item{}#1}} 

\def\listrefs{\vfill\eject\immediate\closeout\rfile
   {{\secfont References}}\bigskip{\frenchspacing%
   \catcode`\@=11\escapechar=` %
   \input refs.tmp\vfill\eject}\nonfrenchspacing}

\def\startrefs#1{\immediate\openout\rfile=refs.tmp\refno=#1}


\global\newcount\figno \global\figno=1
\newwrite\ffile
\def\fig{\the\figno\nfig}
\def\nfig#1{\xdef#1{\the\figno}\ifnum\figno=1\immediate
	\openout\ffile=figs.tmp\fi\global\advance\figno by1\chardef\wfile=\ffile
	\immediate\write\ffile{\medskip\noexpand\item{Fig.\ #1:\ }%
	\figlabel{#1}\pctsign}\findarg}

\def\listfigs{\vfill\eject\immediate\closeout\ffile{\parindent48pt
	\baselineskip16.8pt{{\secfont Figure Captions}}\medskip
	\escapechar=` \input figs.tmp\vfill\eject}}

\def\noblackbox{\overfullrule=0pt}
\def\inv{^{\raise.18ex\hbox{${\scriptscriptstyle -}$}\kern-.06em 1}}
\def\dup{^{\vphantom{1}}}
\def\Dsl{\,\raise.18ex\hbox{/}\mkern-16.2mu D} 
\def\dsl{\raise.18ex\hbox{/}\kern-.68em\partial}
\def\slash#1{\raise.18ex\hbox{/}\kern-.68em #1}
\def\lspace{}
\def\lbspace{}
\def\boxeqn#1{\vcenter{\vbox{\hrule\hbox{\vrule\kern3.6pt\vbox{\kern3.6pt
	\hbox{${\displaystyle #1}$}\kern3.6pt}\kern3.6pt\vrule}\hrule}}}
\def\mbox#1#2{\vcenter{\hrule \hbox{\vrule height#2.4in
	\kern#1.2in \vrule} \hrule}}  
\def\bar{\overline}
\def\e#1{{\rm e}^{\textstyle#1}}
\def\del{\partial}
\def\curly#1{{\hbox{{$\cal #1$}}}}
\def\curlyD{\hbox{{$\cal D$}}}
\def\curlyL{\hbox{{$\cal L$}}}
\def\vev#1{\langle #1 \rangle}
\def\psibar{\overline\psi}
\def\lform{\hbox{$\sqcup$}\llap{\hbox{$\sqcap$}}}
\def\darr#1{\raise1.8ex\hbox{$\leftrightarrow$}\mkern-19.8mu #1}
\def\half{{\textstyle{1\over2}}} 
\def\roughly#1{\ \lower1.5ex\hbox{$\sim$}\mkern-22.8mu #1\,}
\def\MSbar{$\bar{{\rm MS}}$}
\hyphenation{di-men-sion di-men-sion-al di-men-sion-al-ly}

\parindent=0pt
\parskip=5pt


\preprint{
\vbox{
\rightline{CERN-TH.7403/94}
\vskip2pt\rightline{SHEP 94/95-04}
}
}
\vskip -1.5cm

\title{\vbox{\centerline{The Renormalization Group and Two Dimensional}
\vskip2pt\centerline{Multicritical Effective Scalar Field Theory}}}
\vskip -1.5cm
\author{\bf Tim R. Morris}{}{}
\addressline{\it CERN TH-Division}
\addressline{\it CH-1211 Geneva 23}
\addressline{\it Switzerland\myfoot{$^*$}{\rm On Leave from Southampton
University, U.K. (Address after 1/10/94). }}
\addressline{\it }
\vskip -1.5cm

\abstract 
Direct verification of the existence of
an infinite set of
multicritical non-perturbative FPs (Fixed Points) for a single
 scalar field in two dimensions,
 is in practice well outside the capabilities
 of the present standard  approximate non-perturbative methods.
 We apply a derivative expansion of
the exact  RG (Renormalization Group) equations in a form which allows the
corresponding FP equations to appear as non-linear eigenvalue
equations  for
the anomalous scaling dimension $\eta$. At zeroth order, only continuum
limits based on critical sine-Gordon models, are accessible. At second order
in derivatives, we perform a general search over all
$\eta\ge.02$, finding the expected first ten FPs, and {\sl only} these.
For each of these we verify the correct relevant qualitative behaviour, and
compute critical exponents, and the dimensions of up to the first ten lowest
dimension operators. Depending on the quantity, our lowest order approximate
description agrees with CFT (Conformal Field Theory) with an accuracy between
0.2\% and 33\%; this requires however that certain irrelevant operators
that are total derivatives in the CFT are associated with ones that are not
total derivatives in the scalar field theory.

\vskip -1.5cm
\date{\vbox{
{CERN-TH.7403/94}
\vskip2pt{SHEP 94/95-04}
\vskip2pt{hep-th/9410141}
\vskip2pt{October, 1994.}
}
} 
\catcode`@=11 
\def\slash#1{\mathord{\mathpalette\c@ncel#1}}
 \def\c@ncel#1#2{\ooalign{$\hfil#1\mkern1mu/\hfil$\crcr$#1#2$}}
\def\lsim{\mathrel{\mathpalette\@versim<}}
\def\gsim{\mathrel{\mathpalette\@versim>}}
 \def\@versim#1#2{\lower0.2ex\vbox{\baselineskip\z@skip\lineskip\z@skip
       \lineskiplimit\z@\ialign{$\m@th#1\hfil##$\crcr#2\crcr\sim\crcr}}}
\catcode`@=12 
\def\nonp{non-perturbative}
\def\phi{\varphi}
\def\epsilon{\varepsilon}
\def\p{{\bf p}}
\def\P{{\bf P}}
\def\q{{\bf q}}
\def\r{{\bf r}}
\def\x{{\bf x}}
\def\y{{\bf y}}
\def\tr{{\rm tr}}
\def\D{{\cal D}}
\def\ins#1#2#3{\hskip #1cm \hbox{#3}\hskip #2cm}
\def\frac#1#2{{#1\over#2}}

Circumstantial evidence strongly suggests that there exists an infinite set
of multicritical \nonp\ FPs for a single scalar field in two
dimensions, corresponding to the universality classes of multicritical
Ising models, equivalently to the diagonal invariants of the unitary minimal
$(p,p+1)$ conformal models with $p=3,4,\cdots$
\ref\zs{A.B. Zamolodchikov, Yad. Fiz. 44 (1986) 821.}
\ref\car{J.L. Cardy, in ``Phase Transitions and Critical Phenomena'',
vol. 11, C. Domb and J.L. Lebowitz ed. (1987) Academic Press.}, however
direct verification of these
facts is in practice well outside the capabilities of the standard
approximate \nonp\ methods: lattice Monte Carlo, resummations of
weak or strong coupling perturbation theory and the epsilon expansion.
(The impracticableness of the epsilon expansion for higher $p$
is covered in ref.\ref\eps{P.S. Howe and P.C. West, Phys. Lett. B223 (1989)
371\semi M. Bauer, E. Br\'ezin and C. Itzykson (1987),
unpublished.},
implying similar difficulties in weak coupling perturbation theory,
 while lattice methods suffer from difficulties of
locating and accurately computing the multicritical points in the at least
$p-2$  dimensional
bare coupling constant space).
In this letter we demonstrate directly that these multicritical points
do exist -- by an approximation scheme which is reviewed below. Our main
motivation for the present letter is to show that the approximation
scheme, which we presented and applied to three and four dimensional
scalar field theory in refs.\nref\erg{T.R. Morris, Int. J. Mod. Phys. A9
(1994) 2411.}\ref\deriv{T.R. Morris,
Phys. Lett. B329 (1994) 241.}\ref\trunc{T.R. Morris, Phys. Lett.
B334 (1994) 355.}, {\sl is} powerful enough, in fact to {\sl automatically
uncover}, and reliably describe, the expected much richer set of
\nonp\ theories in two dimensions. Many of our
operator dimensions are computed to at least the same accuracy
as that of the most relevant operator -- very much in  contrast to all the
above
approximation methods, suggesting that our equations may also
provide a reliable  description well away from FPs.
These facts are certainly encouraging for our
ambition of developing a reliable and accurate analytic approximation
method of general applicability to \nonp\ quantum field theory\erg--\trunc.

As mentioned in the abstract we use the approximation scheme carried to the
lowest sensible order, to verify that each of the first ten multicritical
points ($p=3,\cdots,12$) have the correct
qualitative behaviour -- e.g. shape of potential, number of relevant
directions,
expected parities under $\phi\leftrightarrow-\phi$
 etc. -- and compute critical exponents, and scaling dimensions
of up to the first ten lowest dimension (integrated) operators.
All of the $\sim100$ quantities agree with CFT to
(sometimes much better than) 33\%, with a weak {\sl improvement} in accuracy
with increasing
multicriticality. However, we find that some integrated irrelevant operators
appear in the scalar field theory spectrum, which can only be identified
with operators which are total derivatives in the CFT.
It appears to be an interesting challenge to understand why these integrated
operators do not vanish.

We start with a review of the method\deriv.
 We wish to formulate quantum field theory
in a way in which it is obvious that our approximations are renormalizable\erg.
We do so by never needing to determine bare quantities, such as a bare action
$S_{\Lambda_0}[\phi]$, or equivalently counterterms.
In this letter we discuss only quantum field theories with no low energy
mass scale. (The generalisation to massive ones is in principle
straightforward). If indeed such theories are independent of the cutoff
$\Lambda_0$
then the (renormalized) theories have scale invariance. As is well known
this is achieved only by assigning scaling dimensions to the operators,
which are different from their na\"\i ve 
dimensions in general. Thus for a single component scalar field $\phi(x)$ in
$D$ (euclidean) dimensions we have that its scaling dimension is
$d_\phi=\half(D-2+\eta)$, where $\eta$ is the anomalous scaling dimension.
For example, this implies  by dimensional analysis that propagators, as a
function of momentum $q$, appear as
\eqn\proscale{<\phi\ \phi>\propto 1/q^{2-\eta}\quad\quad.}
To write the condition that the theories be scale invariant, and to formulate
an efficacious approximation scheme, it is helpful to introduce a (low
energy) scale\foot{In the Wilson sense this is the intermediate cutoff
at which one defines the effective action\ref\wil{K. Wilson and J. Kogut,
Phys. Rep. 12C (1974) 75.}. This connection is developed in ref.\erg.}\
$\Lambda$. Now by dimensions, \proscale\ appears as
\eqn\prolam{<\phi\ \phi>\propto {1\over\Lambda^{2-\eta}}f(q^2/\Lambda^2)}
for some function $f$. We will require that $\Lambda$ is introduced in such
a way that $f$ can be Taylor expanded for small
$q^2/\Lambda^2$, because our approximation will follow from a momentum
expansion (equivalently derivative expansion) by dropping terms beyond
some maximum order. This requirement
implies that the infrared singularity in \proscale\
has been smoothly regulated. Therefore $\Lambda$ is equivalent to a smooth
infrared cutoff. We introduce it into the partition function $Z[J]$ as
$C(q,\Lambda)$, satisfying $C(q,\Lambda)\to 0$ as $q\to 0$, by writing
$S_{\Lambda_0}[\phi]\mapsto S_{\Lambda_0}[\phi]+\half\phi.C^{-1}.\phi$,
where only $C$ depends on $\Lambda$.
{}From this it is straightforward
to write down a differential equation for $Z[J]$ with respect to $\Lambda$.
It turns out however to be helpful to transform this  to a differential
equation
for the Legendre effective action. (One
important reason is because the integrals involved will converge,
with our choice of $C$,  only if the
full self-energy is used -- expanded as a power series to the prescribed
maximum order). In terms of this, after expressing $\phi$, $C$ and $q$ as
dimensionless quantities\foot{$C$ must be chosen to scale correctly\deriv.}\
using $\Lambda$, we find\deriv\
\eqn\scag{\eqalign{
({\partial\over\partial t}&+d_\phi\Delta_\phi+\Delta_\partial
-D)\Gamma[\phi] =\cr
&-\zeta\int_0^\infty\!\!\!\!\!dq\, q^{D-1}
\left({q\over C(q^2)}{\partial C(q^2)\over\partial q} +2- \eta\right)
\left\langle\left[1+C.{\delta^2\Gamma\over\delta\phi\delta\phi}
\right]^{-1}\mkern-23mu(\q,-\q)\right\rangle\ \ .\cr}}
Here $t=\ln(\mu/\Lambda)$ with $\mu$ some arbitrary reference scale.
The angle brackets refer to an average over all directions
of the vector $\q$. $\zeta$ is a normalization factor, introduced
for convenience, by a numerical rescaling.
$\Delta_\phi=\phi.{\delta\over\delta\phi}$ counts the number of fields in a
given vertex, and $\Delta_\partial$ counts the number of derivatives
in a given vertex\deriv.

The requirement of scale invariance is now simply given by
 ${\partial\over\partial t} \Gamma[\phi]=0$.
Substituting this and a derivative expansion
\eqn\gexp{\Gamma[\phi]=\int\! d^Dx\,\{
V(\phi,t)+\half(\partial_\mu\phi)^2K(\phi,t)+(\partial_\mu\phi)^4H_1(\phi,t)
+(\lform\phi)^2H_2(\phi,t)+\cdots\} \ \ ,}
into \scag\ and expanding the RHS (Right Hand Side) to some maximum order
in derivatives yields $n$ coupled second order non-linear ordinary
differential equations for the {\sl fixed point} coefficient functions
 $V(\phi),K(\phi),H_1(\phi),H_2(\phi),\dots$, where
$n$ is the number of undetermined coefficient functions.
As such, we expect at first
sight there to be a $2n$ parameter set of solutions. In fact, generally
there are only a discrete set of possible solutions. Of course only a discrete
set of solutions is generally expected on physical grounds: they correspond
to the possible massless continuum limits (continuous phase transitions) with
the prescribed field content. Mathematically this arises because nearly all
choices of BCs (Boundary Conditions) for the differential equations lead to
solutions
with singular behaviour at some finite real value of the field $\phi$
\deriv\trunc. Another way of seeing why only a discrete set of solutions is
allowed is as follows. We take for simplicity a theory with $Z_2$ symmetry
$\phi\leftrightarrow-\phi$, and assume $d_\phi\ne0$. (See later
for what happens when these conditions are relaxed).
In this case there are $n$ BCs
given by symmetry -- namely that the first derivative of the
coefficient functions ($V'(\phi)$, $K'(\phi)$, etc.) vanish at $\phi=0$,
while a further $n$ BCs are given by the behaviour of the
coefficient functions for large $\phi$ which are determined, up to a
proportionality constant, by dimensional analysis -- assuming that in this
limit
$\Lambda$ can be neglected:
\eqn\sclaw{V(\phi)\propto \phi^{D/d_\phi}, \quad
K(\phi)\propto \phi^{(D-2)/d_\phi -2},\quad \cdots \ins11{as} \phi\to\infty\ .}
Since these $2n$ conditions\foot{If $Z_2$ symmetry is dropped,
\sclaw\ for $\phi\to\pm\infty$ provides all $2n$ conditions.}\
 are imposed on a $2n$ parameter set, we again
expect only a discrete set of solutions.
Actually this is not the whole story because the parameter $\eta$ in \scag\
(with  ${\partial\over\partial t} \Gamma[\phi]=0$) must still be determined.
Generally this is only possible if
there is a reparametrization invariance of the RG
equations. This turns the equations into non-linear eigenvalue equations
for $\eta$. (To see this, use the invariance to fix an
extra condition). Clearly it is important that such an invariance
is preserved by the approximation scheme.
We can do this if we choose $C(q^2)$ to be homogeneous in $q$, i.e.
$C(q^2)\propto q^{2\kappa}$, since then a reparametrization invariance of the
equations exists according to the following (non-physical) scaling dimensions:
\eqn\ssym{\eqalign{
&[\partial_\mu]=[q_\mu]=1, \quad\quad [\phi]=\kappa+D/2,\cr
{\rm hence}\hskip 2cm [V]=&D,\quad  [K]=-2(\kappa+1),\quad
[H_i]=-D-4(\kappa+1) \ \ ,\cr}}
and this is clearly not affected by neglecting higher derivative terms.
The value of $\kappa$ may be determined uniquely, by considerations
of convergence, to be the smallest integer larger than $D/2-1$ \deriv.

The operator spectrum may be determined by linearization of \scag\ about
the FP solutions. By separation of variables the perturbations are of
the form $\delta V(\phi,t)=\epsilon (\mu/\Lambda)^\lambda\, v(\phi)$,
$\delta K(\phi,t)=\epsilon (\mu/\Lambda)^\lambda\, k(\phi)$,
 $\delta H_i(\phi,t)=\epsilon (\mu/\Lambda)^\lambda\, h_i(\phi)$, etc, for some
functions $v$, $k$, $h_i$ etc, where $\epsilon$ is infinitessimal and $\lambda$
appears as an eigenvalue for the linearized equations. In fact $\lambda$
is quantized since we have again $2n+1$ constraints:
$n$ from symmetry ($v'(0)=0$, $k'(0)=0$, $\cdots$ for even eigenfunctions,
or $v(0)=0$, $k(0)=0$, $\cdots$ for odd eigenfunctions), $n$ from dimensional
analysis ($v(\phi)\propto \phi^{(D-\lambda)/d_\phi}$,
$k(\phi)\propto \phi^{(D-2-\lambda)/d_\phi-2}$, $\cdots$ as $\phi\to\infty$,
again providing $\Lambda$ can be ignored in this limit), and one from
a normalization constraint allowed by linearity. $\epsilon\mu^\lambda$
plays the r\^ole of an infinitessimal coupling constant of dimension
$\lambda$, conjugate to an integrated operator of the form
$\gamma[\phi]=\int\! d^D\!x\,\{ v(\phi)
+\half k(\phi)(\partial_\mu\phi)^2+\cdots\}$.
Note that this implies that the operator itself has dimension $D-\lambda$.

In interpreting the operator spectrum it is important to discard any redundant
operators\ref\weg{F.J. Wegner, J. Phys. C7 (1974) 2098.}. These are
operators of the form\deriv\
\eqn\red{\gamma[\phi]=\int\!d^D\!x\, F_\x[\phi]
\,\delta\Gamma[\phi]/\delta\phi(\x)\quad, }
corresponding to reparametrizations of the effective action.
They have no physical
significance, and no well-defined scaling dimension since it
depends on the details of the
RG used\weg. Since we already have $2n+1$ constraints, the requirement \red\
will overconstrain the problem leading to no solutions unless the redundant
operators exist for special reasons (viz. symmetries). We know of two such
operators. One is the operator corresponding to the symmetry \ssym, which
thus has (with some arbitrary normalisation)
$F_\x[\phi]=-x^\mu\partial_\mu\phi(\x)-(\kappa+D/2)\phi(\x)$,
$v(\phi)=D V(\phi)-(\kappa +D/2) \phi V'(\phi)$,
$k(\phi)=-2(\kappa+1)K(\phi)-(\kappa+D/2)\phi K'(\phi)$,
$\cdots$, even parity,
and eigenvalue $\lambda=0$. And the other\weg\ corresponds to the
$\phi$ translation symmetry of the {\sl unscaled} RG
equations. It has $F_\x[\phi]=1$,  $v(\phi)=V'(\phi)$, $k(\phi)
=K'(\phi)$, $\cdots$, and odd parity. By operating with $\int\!d^D\!x\,
\delta/\delta\phi(\x)$ on \scag, it can be seen that, to any order
of the derivative expansion, $\lambda=d_\phi$, as
expected on dimensional grounds.

{}From now on we set $D=2$. This implies $\kappa=1$.  To
$O(\partial^2)$ we drop all derivatives higher than second on the RHS of \scag.
Since the higher FP coefficient functions $H_i,\cdots$,
then satisfy linear equations given by the LHS of \scag\ with
 ${\partial\over\partial t} \Gamma[\phi]=0$, they must vanish, otherwise
they have singularities at $\phi=0$ (for general values of $\eta$) which is
unacceptable. Thus the inverse operator
in \scag\ is the same as that computed in ref.\deriv. Performing the average,
the $q$ integral, choosing  $\zeta=1/4$, and matching both sides of \scag\
we find for $V(\phi,t)$ and $K(\phi,t)$:
\eqna\ii\
$$\eqalignno{&\phantom{\hskip1cm\hbox{and}\hskip1cm}\
{\partial V\over\partial t}+{\eta\over2}\phi V'-2V=
-\left(1-{\eta\over4}\right) P_K(V'') &\ii a\cr
&\hskip1cm\hbox{and}\hskip1cm
{\partial K\over \partial t}+{\eta\over2}\phi K' +\eta K=
\left(1-{\eta\over4}\right)\Biggl\{ P_K(V'') \Biggl[
{V''K''-4V'''K'\over2({V''}^2-4K)}&\ii b\cr
&\ \ \ +{2KV''{V'''}^2+13V''{K'}^2-40KK'V'''\over2({V''}^2-4K)^2}
+10K{KV''{V'''}^2-4KK'V'''+V''{K'}^2\over({V''}^2-4K)^3}\Biggr]\cr
&-{12KK''+11{K'}^2\over24K({V''}^2-4K)}
+{20K'V''V'''-11K{V'''}^2-44{K'}^2\over6({V''}^2-4K)^2}
-10K{K{V'''}^2-K'V''V'''+{K'}^2\over({V''}^2-4K)^3}\Biggr\}\cr}$$
\eqn\pba{\eqalign{
\ins01{where} P_b(a) &=
{\tanh^{-1}\sqrt{1-4b/a^2}\over\sqrt{a^2-4b}}\ins11{if} a^2>4b\cr
 &= {\tan^{-1}\sqrt{4b/a^2-1}\over\sqrt{4b-a^2}}\ \ \ins11{if} a^2<4b\ \ ,
\cr}}
and $\tan^{-1}$ is taken in the range $0\le \tan^{-1}\le \pi$. Eqns. \ii{}\
hold true only if
\eqn\conds{K>0 \ins22{and} V''>-2\sqrt{K}\quad,}
 for otherwise the integrals diverge at unphysical poles.
These conditions are sufficient to ensure that the obvious
physical stability requirements are satisfied.

Consider now the $O(\partial^0)$ case, where the RHS of \ii b\ is also dropped.
Then $\eta=0$, since otherwise
the FP $K$ solution is singular: $K(\phi)\propto 1/\phi^2$.
With $\eta=0$, $K(\phi)$ is not determined. We will here simply
assume that $K(\phi)\equiv1$ is a good approximation. The remaining FP
equation is now $2V(\phi)=P_1(V'')$. Since $P_1(a)$ is a positive
monotonically decreasing function which diverges at $a=-2$, the `potential'
$U(x)=-\int^x\!\! dy\, P_1^{-1}(2y)$ is bounded below
with a single stationary point. If we think of $V$ as
position $x$, and $\phi$ as the time $\tau$, then we see that,
apart from the Gaussian FP  ($V(\phi)\equiv\pi/8$),
$V(\phi)$ has only periodic solutions, as follows from Newton's
equation $d^2x/d\tau^2=-\partial U/\partial x$. These correspond to
the semi-infinite line of critical sine-Gordon models (that is circular
bosons with a radius tunable to any value larger than some minimum radius).
  It would take
us too far from our present purpose however to flesh this out.
Since $d_\phi=0$, eqn. \sclaw\ does not apply; in fact the unscaled
$V(\phi)$ is nowhere
independent of $\Lambda$ and vanishes in the limit $\Lambda\to0$.
(Incidentally, the $d_\phi\phi V'(\phi)$ term in \ii a\  corresponds,
in  the above Newtonian analogy, to a {\sl negative} friction term proportional
to velocity and to the time. Physically it is then easy to understand that this
term is responsible for generic singular behaviour
 and that the careful balancing act $\sclaw$ is necessary if this
is to be avoided.)
\midinsert
\centerline{
\psfig{figure=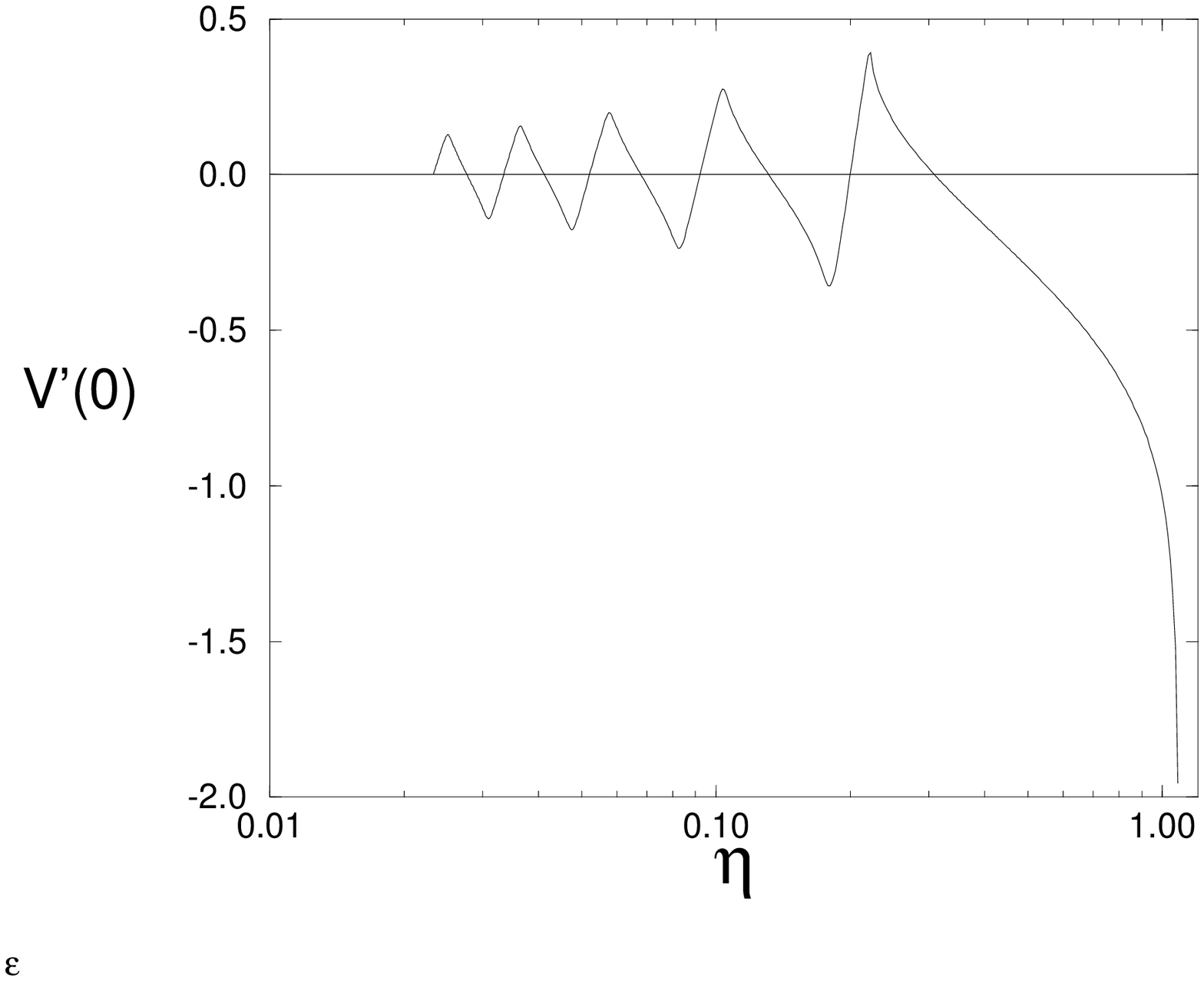,width=4in}}
\vskip 0in
\centerline{\vbox{\baselineskip=14.4pt plus 0.3 pt minus 0.3pt
{\bf Fig.1.} $V'(0)$ versus $\eta$.
Note that no solution was found for $\eta>1.08$.
The $p=3\cdots12$ FPs correspond to the 10 places where $V'(0)$ crosses
the axis.
}}
\endinsert

Returning to the full $O(\partial^2)$ equations, we
substitute  a differentiated \ii a\ into the $V'''$
terms in \ii b\  to turn \ii{}\ manifestly into a pair of
second order differential
equations. We analyse their FPs
 by the
numerical method of relaxation\ref\NR{W.H. Press et al, ``Numerical Recipes ..
The Art of Scientific Computing'', 2nd edition (1992) C.U.P.}.
We can find the approximate $p=3$ (Ising model) solution and its eigenvalue
$\eta=\eta_I$ by using a full set of BCs
and supplying an initial guess (e.g. the $D=3$
Ising solution\deriv), however  finding the other solutions this
way is much harder.
 (More details of the numerics are given in the appendix).
Anyway, we want to demonstrate that for $d_\phi\ne0$, the
eqns.\ii{}\ have no other FPs except the $(p,p+1)$ points. This
is the expected result, by Zamolodchikov's $c$-theorem\ref\zc{A.B.
Zamolodchikov, JETP Lett. 43 (1986) 730.},
if the FP can be the result of flow from a
Landau Ginsburg Lagrangian based about the $c=1$ ultraviolet Gaussian
FP. In fact we will here strictly only
show that no other {\sl connected}
 FPs exist in the following sense:
Starting from our approximate
Ising FP, we relax the BCs by dropping the requirement
$V'(0)=0$ and instead determine a unique solution
 as a function of $\eta$ by using relaxation while
gradually changing $\eta$ away from $\eta_I$.
The result is plotted in fig.1
for all $\eta\ge.02$. Each time $V'(0)$ crosses the axis we store the solution
and use this as a guess for `polishing' by relaxation with the full set of BCs.
(A more conclusive demonstration that there are no other FPs would involve
relaxing e.g. the
$K'(0)$ condition also, and searching a two dimensional parameter space).
Similarly, using relaxed BCs and starting from
 the exactly known operators given earlier or below,
we can determine all\foot{Here we can be
confident that all solutions are connected since we do not expect or
find at any point a
two dimensional linear space of solutions to the relaxed BCs.}\ the
even and odd parity
eigenoperators and eigenvalues at any of the FPs.
Here  one must be careful that
the normalisation condition can be satisfied at any value $\lambda$.
We chose $v'(0)^2+v(0)^2=1$, dropping the requirement $v'(0)=0$
($v(0)=0$) for the even(odd) parity solutions.

Some results are displayed
in the table below.
The dimensions and parities of the two most
relevant operators are not displayed since the  unit operator $v(\phi)=1$,
$k=h_i=\cdots=0$, has dimension 0 and even parity for any FP and
to any order in the derivative expansion, while the
field itself $v(\phi)=\phi$, $k=h_i=\cdots=0$, always has dimension
$d_\phi$ ($=\eta/2$ in this case)
and odd parity. (These follow easily from the vanishing of the RHS of \scag\
linearized about the FP).
The correlation length critical exponent is given by
$\nu=1/\lambda$ where $\lambda$ is the largest even parity eigenvalue,
excluding the unit operator\wil.
All numbers were determined to an accuracy greater than that
shown.\foot{Thus the small dimensions at the bottom of the table correspond
to $\lambda$'s computed to better than 5 significant figures, and agreeing
with CFT to 3 significant figures.}\
Also shown in the table are the corresponding
results expected from CFT\zs\car\ref\revs{See for example the reviews
A.B. Zamolodchikov and Al.B. Zamolodchikov, Sov. Sci. Rev. A. Phys. 10 (1989)
269\semi J.L. Cardy, ``Conformal Invariance and Statistical Mechanics'',
Les Houches Summer School (1988) 169.}.
These follow from the dimensions  $\Delta_{n,m}={[(p+1)n-pm]^2-1\over2p(p+1)}$
and $Z_2$ parities $P_{n,m}=(-)^{(n+1)(p+1)+(m+1)p}$ of the scalar
primary fields
$\Phi_{n,m}(x)$, $\Phi_{1,1}\equiv1$ and $\Phi_{2,2}\equiv\phi$,
where $m,n$ are integers in the range
$1\le n\le p-1$, $1\le m\le p$. In addition there are an infinite number
of Virasoro descendents, however all but $L_{-1}{\tilde L}_{-1}\Phi_{n,m}
\equiv\lform\Phi_{n,m}$ have too high a dimension to correspond to
the $O(\partial^2)$ results in the table. 
Thus $2p-3$ relevant operators are expected ($\Phi_{k,k}$, $k=1,\cdots,p-1$
and $\Phi_{k+1,k}$, $k=1,\cdots,p-2$), of which $p-1$ are even and $p-2$
are odd. We checked that this is true for all our solutions.
In fact, as one can see from the table,
these  CFT values and our results match unambiguously.
As further evidence for the correspondence we note that if the bare Lagrangian
is taken  to be of Landau Ginsburg type with a potential which is an
even polynomial in $\phi$ of degree $2p-2$ \zs\car, then we might expect the
FP Wilson potential to have $p-1$ minima; It does. Examples are shown
in fig.2 and ref.\ref\conf{For graphs of $V$ and $K$
at $p=12$ see T.R. Morris, in Proceedings of XII International Symposium
on Lattice Field Theory, LATTICE '94, Bielefeld, Germany, Eds. F. Karsch et al,
Southampton preprint SHEP 94/95-10, hep-lat/9411053.}.
Similarly, the $v(\phi)$ component of the operators turn out to
have the maximum number of nodes expected (c.f. fig.4).
\midinsert
\centerline{
\psfig{figure=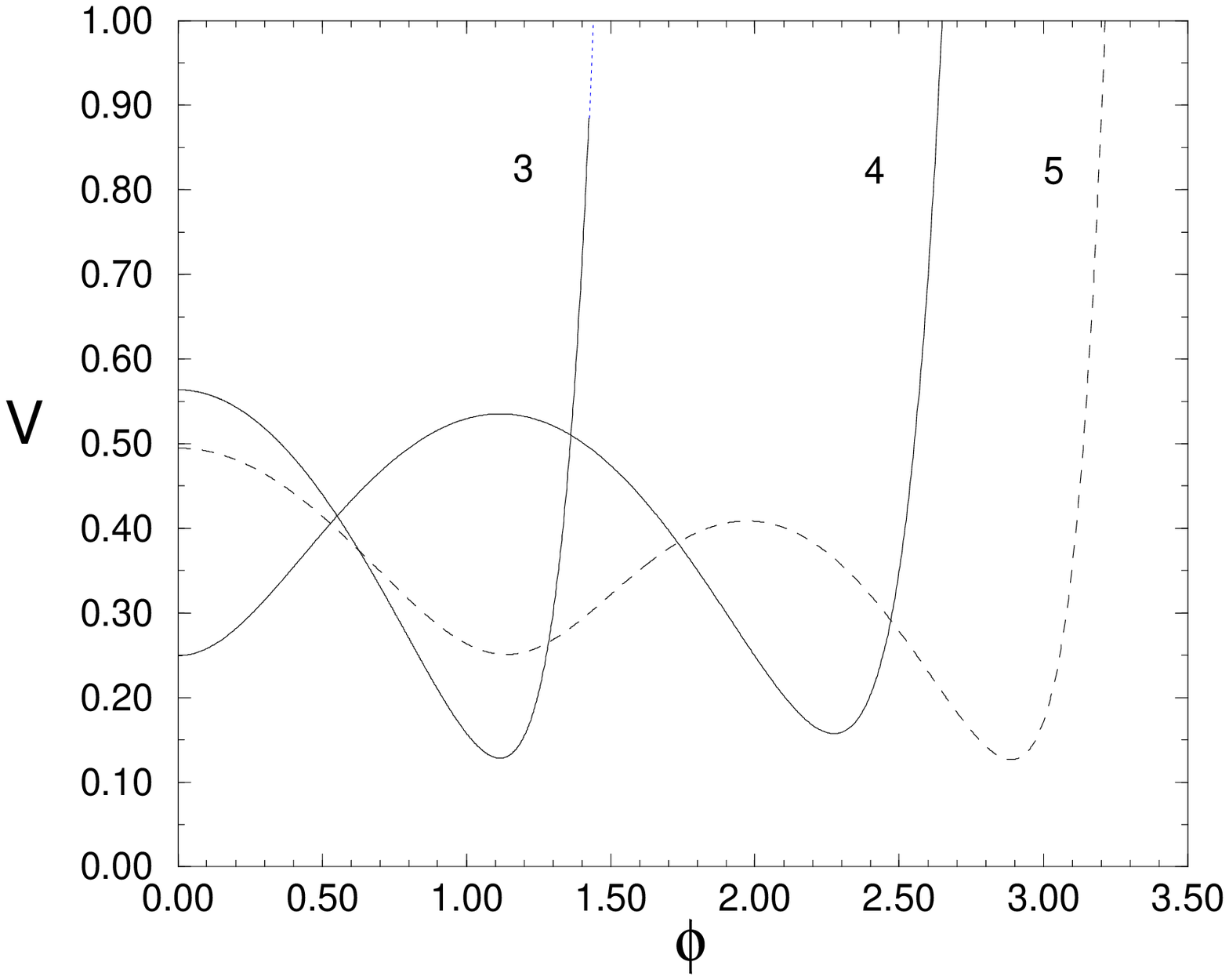,width=4in}}
\vskip 0in
\centerline{\vbox{\baselineskip=14.4pt plus 0.3 pt minus 0.3pt
{\bf Fig.2.} Potentials $V(\phi)$ for the Ising ($p=3$),
tricritical Ising ($p=4$), and quadricritical Ising ($p=5$) fixed points.
}}
\endinsert
\midinsert
\centerline{
\psfig{figure=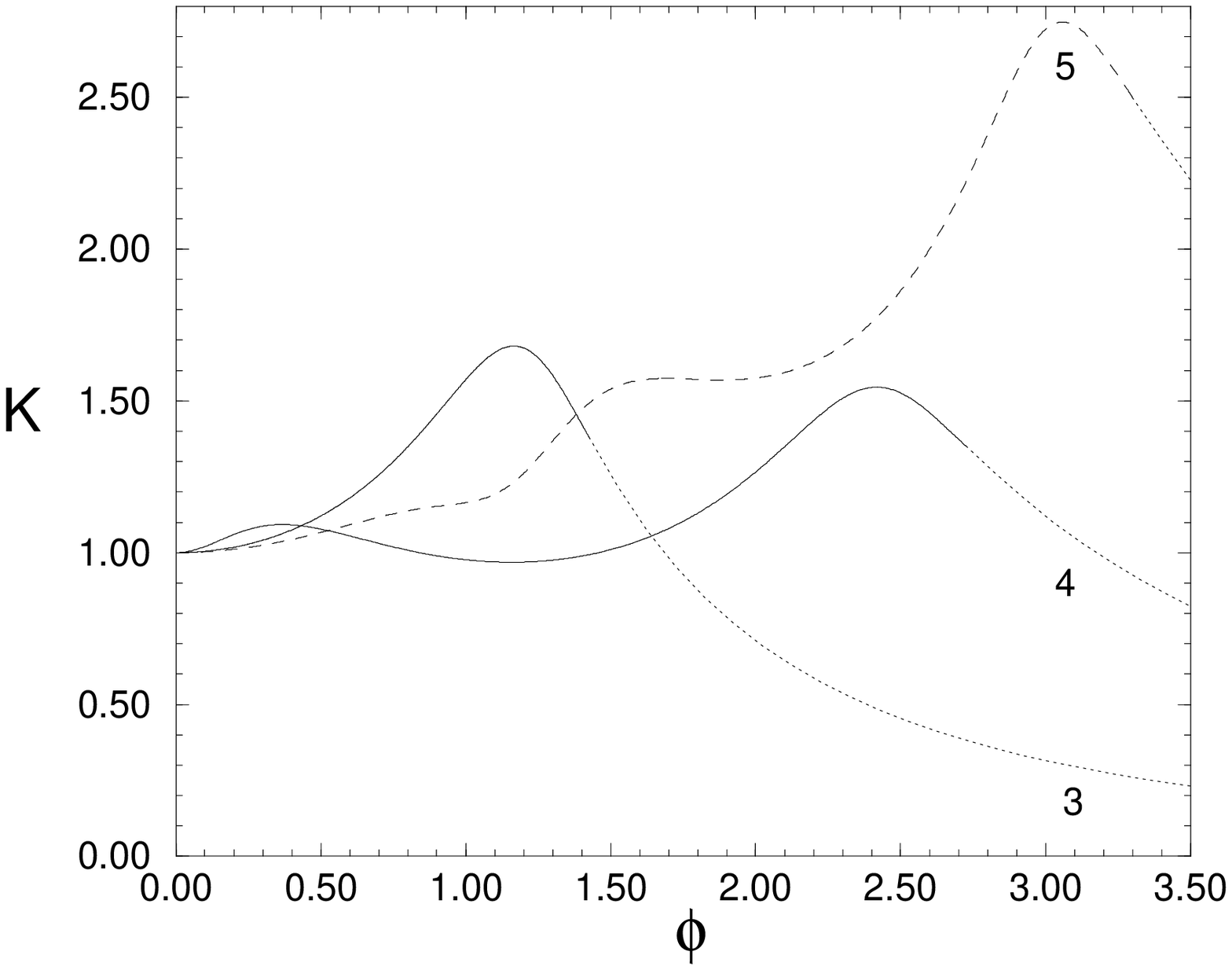,width=4in}}
\vskip 0in
\centerline{\vbox{\baselineskip=14.4pt plus 0.3 pt minus 0.3pt
{\bf Fig.3.} Fixed point
 kinetic factors $K(\phi)$ for $p=3,4,5$.
As in fig.2, the full solution is shown as a full (or for $p=5$ dashed)
line, while the dotted parts of the curve are given by the asymptotic
expansion  (using $A_V$ and $A_K$).
Eqn.\ssym\ has been used to scale the solutions to conventional
normalisation $K(0)=1$,
from the $K''_{asy}(1)=0$ normalisation. (See appendix).
}}
\endinsert
\midinsert
\centerline{
\psfig{figure=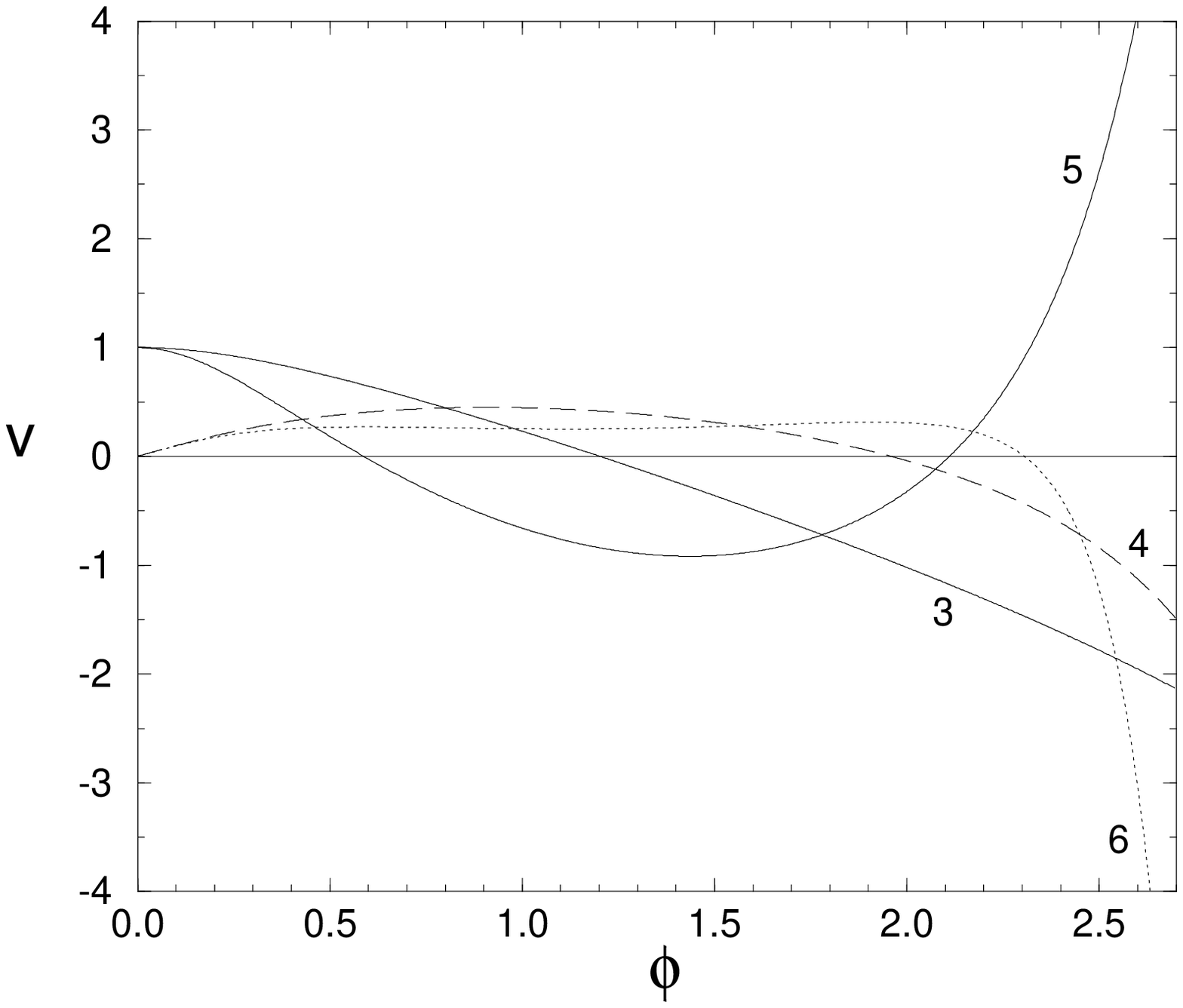,width=4in}}
\vskip -.5cm
\centerline{\vbox{\baselineskip=14.4pt plus 0.3 pt minus 0.3pt
{\bf Fig.4.} The $v(\phi)$ component of the $3^{\rm rd}$ --
$6^{\rm th}$ eigenoperators at the tricritical Ising model fixed point ($p=4$).
As in
figs.1 and 2, the solutions have been scaled to $K(0)=1$ normalisation.
Furthermore the even(odd) operators have been normalised so that $v(0)=1$
($v'(0)=1$). Note
that the $1^{\rm st}$ operator thus has $v(\phi)\equiv1$, and the $2^{\rm nd}$
has
$v(\phi)\equiv\phi$. The $3^{\rm rd}$ -- $5^{\rm th}$ are identified at the
bare level with the powers $\phi^2$ -- $\phi^4$ in refs.\zs\car (plus
lower dimension counterterms of the same $Z_2$ symmetry),
while the $6^{\rm th}$ is one of the ``puzzling operators''.
}}
\endinsert

We do not find all
the irrelevant operators expected from CFT, which is why there are blank
entries in the table. For each $p$ we obtain a cluster of solutions and
then a gap in dimensions,  much larger than expected from CFT.
 We report the $p=3,4,5$ cases in detail.
For $p=3$, we find the first operator, other than those in the table, is
an even operator with
 $\Delta=20\pm2$. (The estimated error is due to truncation in the asymptotic
BCs). In the $p=4$ spectrum we find only those operators in the table,
in particular no sign of $L_{-2}{\tilde L}_{-2}\Phi_{2,2}$ (dimension and
parity $\Delta^P=4{3\over
40}^-$). In $p=5$, we find an $11^{\rm th}$ and  a $12^{\rm th}$
operator with $\Delta^P=$ $2.61^+$ and $2.80^-$ respectively,
and then nothing else
out to at least $\Delta=8$, and in particular
the primary $\Phi_{4,1}$ ($\Delta^P=6^+$) is missing.
However, high dimension
operators can correspond to operators with many derivatives, and therefore
may not appear at our level of approximation.

More puzzling are the integrated operators that are there, but (apparently)
should not be because
they only match total derivative operators in CFT (given as boldface
entries in the table).
Thus at $p=3$, the
$4^{\rm th}$ operator's $\Delta^P$ matches well with
$\lform \Phi_{2,2}$, and not with any
other operator -- in particular the first
odd operator that is not a total derivative has $\Delta=6{1\over8}$.
Similarly all the other irrelevant dimensions mentioned, match
 $\lform$ on the relevant CFT operators, with
the $11^{\rm th}$($12^{\rm th}$) operators in $p=5$ matching
the values $2{4\over5}^+$($3{1\over20}^-$).
 Only for the $7^{\rm th}$ operator in
$p=4$, and the $11^{\rm th}$ and $12^{\rm th}$ in $p=5$, would alternative
assignments be possible within the accuracy suggested by the table:
to $\Phi_{3,1}$ ($3^+$), $\Phi_{3,1}$ ($2{4\over5}^+$),
and $\Phi_{4,2}$ ($3{1\over4}^+$) respectively.
\vskip15pt plus4pt minus4pt
\centerline{
\vbox{\offinterlineskip\hrule\halign{&\vrule#&\strut\ #\ \hfil\cr
&\hfil $p$&&\hfil$\eta$&&\hfil$\nu$&&\hfil$3^{\rm rd}$&&\hfil$4^{\rm th}$&&
\hfil$5^{\rm th}$&&\hfil$6^{\rm th}$&&\hfil$7^{\rm th}$&&
\hfil$8^{\rm th}$&&\hfil$9^{\rm th}$&&\hfil$10^{\rm th}$&\cr
\noalign{\hrule}
   &\hfil3&&.309&&.863&&.841$^+$&&2.61$^-$&& &&
             && && && && &\cr
     & &&.25&&1&&1$^+$&&{\bf 2.13}$^-$&& &&
             && && && && &\cr \noalign{\hrule}
     &\hfil4&&.200&&.566&&.234$^+$&&.732$^-$&&1.09$^+$&&
        2.11$^-$&&2.44$^+$&&2.71$^-$&& && &\cr
       & &&.15&&.556&&.2$^+$&&.875$^-$&&1.2$^+$&&
   {\bf 2.08$^-$}&&{\bf 2.2}$^+$&&{\bf 2.88}$^-$&& && &\cr \noalign{\hrule}
     &\hfil5&&.131&&.545&&.166$^+$&&.287$^-$&&.681$^+$&&
     .953$^-$&&1.26$^+$&&2.11$^-$&&2.16$^+$&&2.38$^-$&\cr
      & &&.1&&.536&&.133$^+$&&.25$^-$&&.8$^+$&&
1.05$^-$&&1.33$^+$&&{\bf 2.05}$^-$&&{\bf 2.13}$^+$&&{\bf 2.25}$^-$&\cr
  \noalign{\hrule}
    &\hfil6&&.0920&&.531&&.117$^+$&&.213$^-$&&.323$^+$&&
     .650$^-$&&.865$^+$&&1.11$^-$&&1.37$^+$&&2.08$^-$&\cr
   & &&.0714&&.525&&.0952$^+$&&.179$^-$&&.286$^+$&&
  .75$^-$&&.952$^+$&&1.18$^-$&&1.43$^+$&&{\bf 2.04}$^-$&\cr \noalign{\hrule}
    &\hfil7&&.0679&&.523&&.0868$^+$&&.159$^-$&&.249$^+$&&
     .348$^-$&&.629$^+$&&.806$^-$&&1.01$^+$&&1.22$^-$&\cr
    & &&.0536&&.519&&.0714$^+$&&.134$^-$&&.214$^+$&&
     .313$^-$&&.714$^+$&&.884$^-$&&1.07$^+$&&1.28$^-$&\cr \noalign{\hrule}
    &\hfil8&&.0521&&.517&&.0667$^+$&&.123$^-$&&.193$^+$&&
     .277$^-$&&.368$^+$&&.613$^-$&&.764$^+$&&.933$^-$&\cr
    & &&.0417&&.514&&.0556$^+$&&.104$^-$&&.167$^+$&&
     .243$^-$&&.333$^+$&&.688$^-$&&.833$^+$&&.993$^-$&\cr \noalign{\hrule}
   &\hfil9&&.0412&&.514&&.0529$^+$&&.0972$^-$&&.154$^+$&&
     .221$^-$&&.299$^+$&&.383$^-$&&.601$^+$&&.733$^-$&\cr
   & &&.0333&&.511&&.0444$^+$&&.0833$^-$&&.133$^+$&&
     .194$^-$&&.267$^+$&&.350$^-$&&.667$^+$&&.794$^-$&\cr \noalign{\hrule}
   &10&&.0334&&.511&&.0429$^+$&&.0790$^-$&&.125$^+$&&
     .180$^-$&&.245$^+$&&.317$^-$&&.395$^+$&&.592$^-$&\cr
   & &&.0273&&.509&&.0364$^+$&&.0682$^-$&&.109$^+$&&
     .159$^-$&&.218$^+$&&.286$^-$&&.364$^+$&&.650$^-$&\cr \noalign{\hrule}
   &11&&.0277&&.509&&.0355$^+$&&.0654$^-$&&.103$^+$&&
     .150$^-$&&.203$^+$&&.265$^-$&&.332$^+$&&.405$^-$&\cr
  & &&.0227&&.508&&.0303$^+$&&.0568$^-$&&.0909$^+$&&
     .133$^-$&&.182$^+$&&.239$^-$&&.303$^+$&&.375$^-$&\cr \noalign{\hrule}
  &12&&.0233&&.508&&.0299$^+$&&.0550$^-$&&.0870$^+$&&
     .126$^-$&&.172$^+$&&.224$^-$&&.282$^+$&&.345$^-$&\cr
  & &&.0192&&.506&&.0256$^+$&&.0481$^-$&&.0769$^+$&&
     .112$^-$&&.154$^+$&&.202$^-$&&.256$^+$&&.317$^-$&\cr
}\hrule}}
\vskip10pt plus3pt minus3pt
{\vbox{
\baselineskip=14.4pt plus 0.3 pt minus 0.3pt
Table 1.  $\eta$,$\nu$, and the $\Delta^P$'s of the 10 lowest
dimension operators (ordered by increasing dimension),
 for the first 10 multicritical points.
The $O(\partial^2)$ answer is shown in the first row
and the associated exact CFT result in the second row.
Worst determined number: $\eta$ for $p=4$ (33\% off). Best determined number:
$\nu$ for $p=12$ (0.2\% off). $\eta$'s accuracy gradually
improves from $p=4$ to $p=12$.
$\nu$ is worst determined at $p=3$ (13\%)
after which all are determined to error less than 2\% and decreasing with
increasing $p$. The worst determined operator dimension is the $3^{\rm rd}$
at $p=5$
(25\%) after which errors decrease with increasing $p$ and/or increasing
dimension. 
The dimensions of the
first half of the relevant operators are always
overestimated while those of the second half are always underestimated.}}

These puzzling solutions could be discarded if they were
redundant, but, as argued below \red,  this requirement overconstrains.
  Indeed, assuming that the
most general (Poincar\'e invariant) redundant operator \red\ is of the form
$F_\x[\phi]=\alpha x^\mu\partial_\mu\phi(\x)+f\left(\phi(\x)\right)+
\cdots$, where $\alpha$ is a constant, $f$ some differentiable
 function, and the rest results
in terms only of
$O(\partial^2)$ or greater in $\gamma[\phi]$, then it is easy to show that
(for any dimension $D$) $\gamma[\phi]$ is a redundant operator only if
$v(\phi_1)V(\phi_2)=V(\phi_1)v(\phi_2)$ for all pairs of stationary points
$\phi_1,\phi_2$ of $V$.
These relations are not even approximately satisfied by our puzzling operators.
Finally, using the cases $p=3,4,5$, the `best fit',
before the large gap in dimensions,  can
be summarised as follows: All the relevant operators
appear, together with a set of irrelevant states whose $\Delta^P$'s match
$\lform$ acting on all these operators but two: the first
($\lform\Phi_{1,1}\equiv0$) and the last ($\lform\Phi_{p-1,p-2}$  is
missing). Note that the $O(\partial^2)$ spectrum has an {\sl unbroken}
oscillating pattern of $Z_2$ parities ($+-+-\cdots+-$), which would have
been broken for $\Delta>2$ had it also (or only) uncovered the
irrelevant primaries here.
\midinsert
\centerline{
\psfig{figure=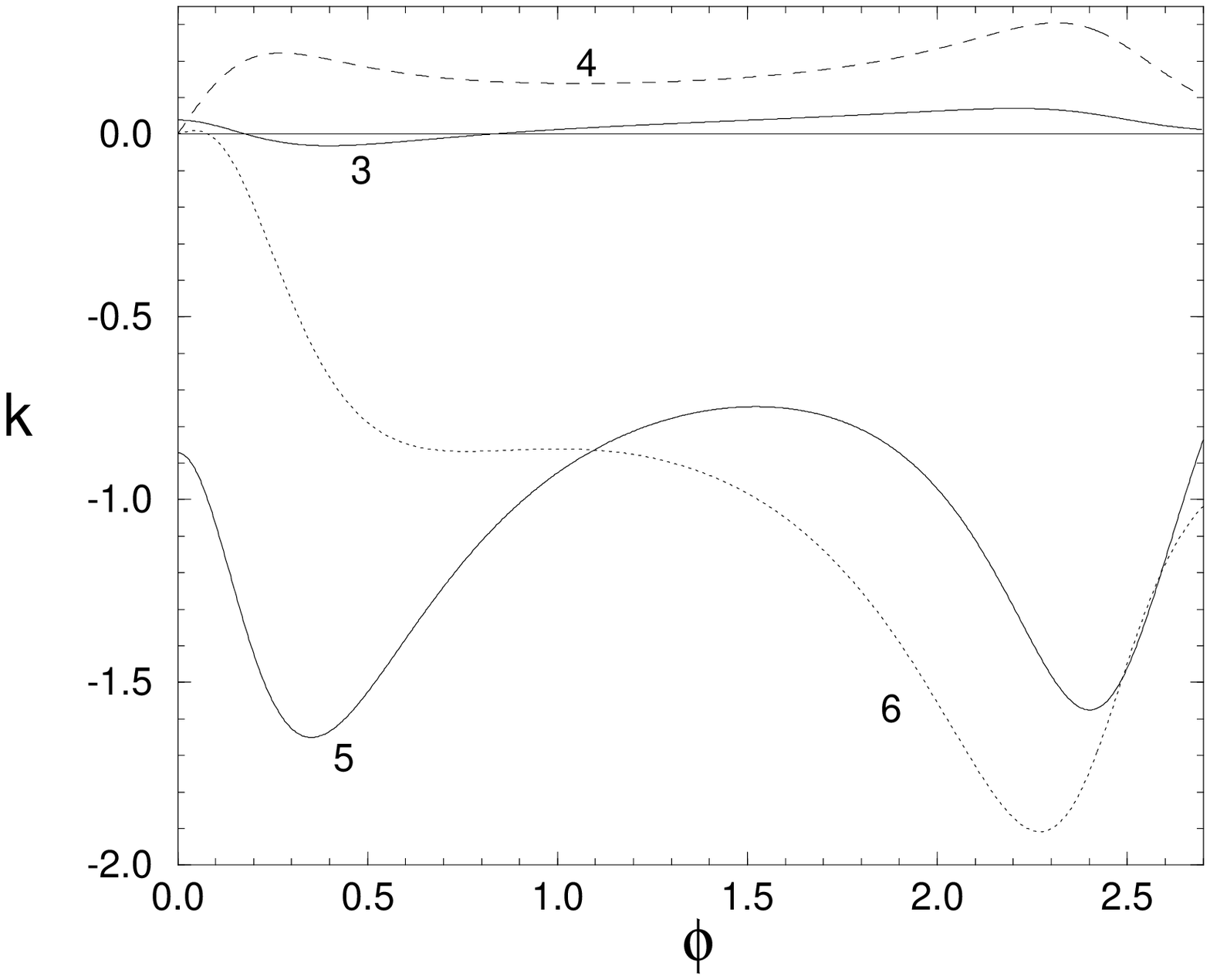,width=4in}}
\vskip 0in
\centerline{\vbox{\baselineskip=14.4pt plus 0.3 pt minus 0.3pt
{\bf Fig.5.} The $k(\phi)$ component of the $3^{\rm rd}$ --
$6^{\rm th}$ eigenoperators at the tricritical Ising model fixed point ($p=4$).
Normalisations are those of fig.4. Recall that
the $1^{\rm st}$ and $2^{\rm nd}$ operator have $k(\phi)\equiv0$.
}}
\endinsert

We briefly mention directions for further research. Evidently,
taking the derivative expansion to higher order would be informative.
We should, for example, begin to see approximate degeneracies in the irrelevant
part of the spectrum.
 Expanding the flow
equations to second order in the perturbations would allow computation of
the operator product (fusion) coefficients, while
non-scalar couplings would allow
access to operators with non-zero spin and e.g. an investigation of
the approximately satisfied Virasoro symmetry.
The full partial differential equations \ii{}, analyzed
e.g. again by relaxation,
can describe
for example flow (crossover) from one FP to another,
using the present analysis to provide the BCs; in particular one
can investigate this way the form of the bare Landau Ginsburg Lagrangian.
 Finally it would be interesting to consider, by relaxation,
 `flow' of the FPs
with (general real) dimension $D$. All of the above suggestions would
help in resolving the reason for our ``puzzling operators''.

{\bf Note added in proof.}

A natural interpretation of our puzzling operators would be
that they are examples of ``shadow operators''
\ref\shad{B. Schroer, Nucl. Phys. B295[FS21] (1988) 586.}. (I thank T.
Hollowood for bringing this work to my attention).
These are operators that exist in the continuum theory but
completely decouple in the conformal
invariant limit. If so, we should expect to find
that, in our approximation, their fusion coefficients with CFT operators
are very small.

\acknowledgements
It is a pleasure to thank John Cardy, Tim Hollowood and {\sl especially} Michel
Bauer for their interest, and advice on CFT and the $\epsilon$ expansion,
and to thank
Martin Hasenbusch for a discussion about Monte Carlo methods.

\appendix{A}{Numerics.}
We sketch some of the
less obvious points involved in the numerical
solution of \ii{}\ at, and linearized about, the FPs.
Note first the reasons for relaxation:
 it is not necessary to cast the eqns \ii{}\
in the form $V''(\phi)=\cdots$ and $K''(\phi)=\cdots$; two point
BCs and eigenvalue problems are easily incorporated; and
it is efficient in the sort of searches illustrated
in fig.1, since the previous solution is a
good guess for the next\NR. It can also cope with stiffness\NR,
which is there because small
perturbations from a solution result in
singularities at some finite $\phi$.
Conditions \conds\ can be violated in the iterands.
If this happened it worked best to back track
and shorten the jumps
 between iterations. It is
necessary to develop the asymptotic BCs \sclaw\
at least to next-to-leading order, introducing explicitly
proportionality constants, say $A_V$ and $A_K$.
This is because we set these
BCs at some finite point $\phi=\phi_{ASY}$,
and imposing just \sclaw\ would imply, by \ii{},
singular values for the
second derivatives. In fact we developed the asymptotics yet one
further order so as to check that the truncation error was well
under control. $\phi_{ASY}$ must lie within
a window bounded below by points where the asymptotic series is not
a sufficiently good approximation and above by points where
roundoff error (due to increasing stiffness)
prevents the relaxation program from converging. This window rapidly shrinks
with increasing $p$ because both effects (we believe) depend on the size of
$\sim\phi^{8/\eta}$. We found
empirically that the upper bound is just above the last turning point
in $K$. This turning point always exists, so
we ensure we are
automatically inside this window by setting $\phi_{ASY}=1$ and
normalising with $K_{asy}''(1)=0$, where  $K_{asy}(\phi)$ is the asymptotic
expansion of $K$.  
We checked that
our results are completely insensitive to moderate
changes in $\phi_{ASY}\lsim1$. We factor out the rapid increase in $V$
in the asymptotic region \sclaw, by writing $V(\phi)=
[V(0)+A_V \phi^{2r}] {\tilde V}(\phi)$, where the integer $r\approx 2/\eta$,
the multiplying factor being
differentiated analytically. Numerical errors are then under control
even for $p=12$ where asymptotically $V\sim\phi^{200}$.
For the eigenvalues $\lambda$, it was
enough to substitute the FP solutions into the linearized
equations, using the same mesh, imposing
the leading asymptotic behaviour  on $v,k$ at $\phi=\phi_{asy}$.
In the positive[negative] parity search,
we recorded those cases where $v(1)$ changes sign but $v'(0)$ [$v(0)$]
does not, since sometimes these are due to passing over eigenvalues, however
cases where both $v'(0)$ [$v(0)$] and  $v(1)$ changed sign were due
to $k(0)$ [$k'(0)$] passing through zero. The most
rapid asymptotic behaviour was factored out of both $v$ and $k$.
For the search we
chose  $\phi_{asy}=1$, but for `polishing' it was often
necessary to choose $\phi_{asy}<1$ for
convergence. We required as much
convergence as possible before choosing $\phi_{asy}$
just inside the remaining wild oscillations, checking insensitivity to
moderate reductions in this value.
Finally  next to leading order $v$ and $k$ asymptotics were used
as a check particularly that no solutions were spurious.

\listrefs
\end